\documentstyle[preprint,aps,epsf,floats]{revtex}

\begin{document}

\vskip 1.5 cm

\preprint{\tighten \vbox{\hbox{} }}

\title{Observables in the Decays of $B$ to Two Vector Mesons}
\author{Cheng-Wei Chiang\footnote{E-mail:
chengwei@andrew.cmu.edu} and Lincoln Wolfenstein\footnote{E-mail:
lincoln@cmuhep2.phys.cmu.edu}}

\address{
Department of Physics,
Carnegie Mellon University,
Pittsburgh, Pennsylvania 15213}

\maketitle

{\tighten
\begin{abstract}
In general there are nine observables in the decay of a $B$ meson to
two vector mesons defined in terms of polarization correlations of
these mesons.  Only six of these can be detected via the subsequent
decay angular distributions because of parity conservation in those
decays.  The remaining three require the measurement of the spin
polarization of one of the decay products.
\end{abstract}
}


\newpage


The decay of the $B$ meson into two vector mesons $B \to V_1\,+\,V_2$,
such as $B \to \rho\,+\,\Psi$ or $B \to \rho\,+\,K^*$, has been
calculated in many models
\cite{DDLR96,DDF98,KP92PRD,KP92PLB,KPS94,KMP92,F99,LSS99,CT99}.  Here
we are concerned with observables from a model-independent viewpoint.
We limit this discussion to the case of $B^{\pm}$ decays or $B^0$
($\bar B^0$) in the absence of $B^0-\bar B^0$ mixing effects.

To take the advantage of extracting the $CP$-odd and $CP$-even or
$T$-odd and $T$-even components more easily, the angular distribution
is often written in the linear polarization (or transversity) basis.
Let us define the amplitude of $B \to V_1 V_2$ in the rest frame of
$V_1$.  According to their polarization combinations, the amplitude
can be decomposed into \cite{DDLR96}
\begin{equation}
\label{ampl}
A(B \to V_1 V_2) = 
   A_0\,{\epsilon}^{*L}_{V_1} {\epsilon}^{*L}_{V_2}
 - \frac{A_{\|}}{\sqrt{2}}\,
     \vec {\epsilon^*_{V_1}}^T \cdot \vec {\epsilon^*_{V_2}}^T
 - i \frac{A_{\perp}}{\sqrt{2}}\,
     \vec {\epsilon^*_{V_1}} \times \vec {\epsilon^*_{V_2}}
     \cdot \hat{\bf p}, 
\end{equation}
and similarly for $\bar B \to \bar V_1 \bar V_2$.  In
Eq.~(\ref{ampl}), $\vec {\epsilon_{V_1}}$ and $\vec {\epsilon_{V_2}}$
are the unit polarization vectors of $V_1$ and $V_2$, respectively.
$\hat{\bf p}$ is the unit vector along the direction of motion of
$V_2$ in the rest frame of $V_1$, $\epsilon_{V_i}^{*L} \equiv \vec
{\epsilon^*_{V_i}} \cdot \hat{\bf p}$ and $\vec {\epsilon^*_{V_i}}^T =
\vec {\epsilon^*_{V_i}} - \epsilon_{V_i}^{*L} \hat{\bf p}$.  It is
easy to see that $A_{\perp}$ is odd under the parity transformation
because of the appearance of $\vec {\epsilon^*_{V_1}} \times \vec
{\epsilon^*_{V_2}} \cdot \hat{\bf p}$, whereas $A_0$ and $A_{\|}$ are
even.

For $B$ decays, the square of the amplitude $A^*A$ should determine 9
observables proportional to products of the three transversity
amplitudes.  We can choose these as given by
\begin{eqnarray}
\label{bilinears}
K_1 = |A_0|^2, & \qquad
K_4 = Re \left[ A_0^*\,A_{\|} \right], & \qquad
L_4 = Im \left[ A_0^*\,A_{\|} \right], \nonumber \\
K_2 = |A_{\|}|^2, & \qquad
K_5 = Im \left[ A_0^*\,A_{\perp} \right], & \qquad
L_5 = Re \left[ A_0^*\,A_{\perp} \right], \\
K_3 = |A_{\perp}|^2, & \qquad
K_6 = Im \left[ A_{\|}^*\,A_{\perp} \right], & \qquad
L_6 = Re \left[ A_{\|}^*\,A_{\perp} \right]. \nonumber
\end{eqnarray}
Then
\begin{eqnarray}
A^*A &=& 
   K_1\,X_1\,+\,K_2\,X_2\,
+\,K_3\,X_3\,+\,K_4\,X_4\,
+\,L_5\,X_5\,+\,L_6\,X_6\, \nonumber \\
&& \qquad
+\,L_4\,Y_4\,+\,K_5\,Y_5\,
+\,K_6\,Y_6 \nonumber
\end{eqnarray}
The observables $X_i$ and $Y_i$ represent polarizations or
polarization correlations of the final vector mesons depending on the
vector $\hat{\bf p}$.  They are given explicitly in the Appendix.

The polarization state of a spin-1 particle is given \cite{L55} in
terms of the spin ${\bf S}_i$ and a second rank traceless tensor ${\bf
T}_{ij}$.  Sometimes the tensor polarization is referred to as
alignment.  The observables can be classified according to their
properties with respect to parity $P$ and motion reversal $T$.  By
``motion reversal'' \cite{Sachs87} is meant the reversal of all spins
and momenta; a nonzero value of a $T$-odd observable signifies
time-reversal violation when there are no final state interactions.
We find:
\begin{center}
\vspace{6pt}
\begin{tabular}{||c|c|c||} \hline\hline
 & P & T \\
\hline
$X_1$ to $X_4$ & even & even \\ \hline
$Y_5$, $Y_6$   & odd  & odd \\ \hline
$Y_4$          & even & odd \\ \hline
$X_5$, $X_6$   & odd  & even \\ \hline\hline
\end{tabular} \vspace{6pt}
\end{center}
The terms which have opposite behaviour under $P$ and $T$ are
necessarily proportional to the spin polarization ${\bf S}_1$ or ${\bf
S}_2$.

The polarization state of $V_i$ is analyzed via its subsequent decay.
If this is a strong or electromagnetic decay into two particles whose
angular distribution is measured, then it is impossible to detect the
spin polarization ${\bf S}_i$ and only ${\bf T}_{ij}$ can be detected.
This is a consequence of parity conservation since the final decay
cannot depend on $\vec{\bf S}\cdot\vec{\bf k}$, where $\vec{\bf k}$ is
the relative momentum of the decay products.  As a result, one cannot
determine $L_4$, $L_5$, and $L_6$ in this way.


In general, the angular distribution of the decay in the transversity
basis can be written as
\begin{equation}
\label{angdist}
\frac{d^3\Gamma}{d\cos\theta_1 d\cos\theta_2 d\phi} =
\sum_{i=1}^6 K_i\,f_i(\theta_1,\theta_2,\phi),
\end{equation}
where $K_i$'s and $L_i$'s are the amplitude bilinears that contain the
dynamics and in the case of $B^0$ would evolve with time\footnote{The
time evolution effects will be addressed in a separate publication
\cite{CL99}.}, and $f_i(\theta_1,\theta_2,\phi)$ are the corresponding
angular distribution functions which are orthogonal to one another.
Here $\theta_1$ (or $\theta_2$) denotes the polar angle of one of the
$V_1$ (or $V_2$) decay products measured in the rest frame of $V_1$
(or $V_2$) relative to the motion of $V_1$ (or $V_2$) in the rest
frame of the $B$ meson, and $\phi$ is the angle subtended by the two
planes formed by decay products of $V_1$ and $V_2$, respectively.

For the case in which the decays of $V_1$ and $V_2$ are both into two
pseudoscalar mesons, one can immediately translate the tensor
correlations into angular distributions as shown in the Appendix.  The
resulting normalized angular distribution of the decays $B \to V_1
(\to P_1 P_1^{\prime}) \, V_2 (\to P_2 P_2^{\prime})$, where
$P_1^{({\prime})}$ and $P_2^{({\prime})}$ denote pseudoscalar mesons,
is:
\begin{eqnarray}
\label{BtoVPPVPP}
\frac{1}{\Gamma_0} \frac{d^3\Gamma}{d\cos\theta_1 d\cos\theta_2 d\phi}
& = & 
\frac{9}{8 \pi \Gamma_0} \biggl \{
  K_1 \cos^2\theta_1 \cos^2\theta_2
+ \frac{K_2}{2} 
       \sin^2\theta_1 \sin^2\theta_2 \cos^2\phi 
\biggr. \nonumber \\
&& 
+ \frac{K_3}{2} 
       \sin^2\theta_1 \sin^2\theta_2 \sin^2\phi
+ \frac{K_4}{2 \sqrt{2}}
       \sin 2\theta_1 \sin 2\theta_2 \cos\phi \\
&&
\biggl.
- \frac{K_5}{2 \sqrt{2}}
       \sin 2\theta_1 \sin 2\theta_2 \sin \phi
- \frac{K_6}{2}
       \sin^2\theta_1 \sin^2\theta_2 \sin 2\phi
\biggr \}. \nonumber
\end{eqnarray}
Here $\theta_1$ ($\theta_2$) is the angle between the $P_1$ ($P_2$)
three-momentum vector in the $V_1 ( V_2)$ rest frame and the $V_1$
($V_2$) three-momentum vector defined in the $B$ rest frame, and
$\phi$ is the angle between the normals to the planes defined by $P_1
P_1^\prime$ and $P_2 P_2^\prime$, in the $B$ rest frame.  An example
is the decay $B^- \to K^{*-} \rho^0 \to \left(K\pi\right)^-
\left(\pi^+\pi^-\right)$.


In order to determine $L_4$, $L_5$, and $L_6$ it is necessary to
measure the spin polarization of a decay product.  For example, if
$V_2$ decays into $\mu^+\,\mu^-$ and $V_1$ decays into
$P_1\,P_1^{\prime}$ then, as shown in the Appendix, the polarization
of either muon is given by
\begin{equation}
\label{muonpolar}
  \frac1{\sqrt{2}} \sin 2\theta_1 \sin \theta_2 
  \left( L_4 \sin\phi - L_5 \cos\phi \right)
+ \frac{L_6}{2}
     \sin^2\theta_1 \cos\theta_2,
\end{equation}
divided by the sum of the first six terms in Eq.~(\ref{BtoPPLL}),
where $\theta_1$ ($\theta_2$) is, as before, the angle between the
$P_1$ ($\mu^-$) three-momentum vector in the $V_1$ ($V_2$) rest frame
and the $V_1$ ($V_2$) three-momentum vector defined in the $B$ rest
frame, and $\phi$ is the angle between the normals to the planes
defined by $P_1\,P_1^{\prime}$ and $\mu^+\,\mu^-$ in the $B$ rest
frame.  For the case of $L_6$ it is only necessary to measure the
polarization of one muon independent of observing the decay of $V_1$.
The maximum value of the polarizations is $1$ for each of the terms
proportional to $L_4$, $L_5$, or $L_6$.

For the case in which $V_1$ and $V_2$ are both $CP$ eigenstates, the
amplitude given by $A_{\perp}$ corresponds to a final state with an
opposite $CP$ eigenvalue from the other two.  This by itself has
nothing to do with $CP$ violation, but it leads to the possibility
that the coefficients $K_5$, $K_6$ involving $A_{\perp}$ may be good
places to look for $CP$ violation.

In particular, in the absence of final state interaction (FSI) it
follows from $CPT$ invariance that only ``$T$-odd'' terms will be odd
under $CP$, so that
\begin{eqnarray}
&& K_i=\bar K_i,\, {\rm for}\; i=1,2,3,4; \nonumber \\
&& K_i=-\bar K_i,\, {\rm for}\; i=5,6; \\
&& L_5=\bar L_5,\; L_6=\bar L_6,\; L_4=-\bar L_4. \nonumber
\end{eqnarray}
and the only signals of $CP$ violation are $L_4$, $K_5$, and $K_6$.
In fact, in the absence of FSI a nonzero value of $L_4$ or $K_{5,6}$
by itself is a signal of time reversal violation.  On the other hand,
in the absence of $CP$ violation (as expected in decays like $B \to
\Psi K^*$) a nonzero value of $K_{5,6}$ is a signal of significant
final state phases.


In order to have $CP$ violation there must be two contributions with
different weak phase factors; we label these as $T_{\eta}$ and
$P_{\eta}$ where $\eta=0, \|, \perp$.  Then each of the three
amplitudes entering Eq.~(\ref{ampl}) has the form
\begin{equation}
\label{decaymatrix}
A_{\eta}
 = e^{i \theta_{\eta}}(T_{\eta}+P_{\eta} \, e^{i \phi_w}
   e^{i \delta_{\eta}}).
\end{equation}
where $\theta_{\eta}$, $\delta_{\eta}$ are strong phases and $\phi_w$
is the relative weak phase between the tree and penguin contributions.
We obtain $\bar A_{\eta}$ by changing $\phi_w$ to $-\phi_w$.  There
are in general 12 parameters: $T_{\eta}$, $P_{\eta}$, $\delta_{\eta}$,
$\phi_w$, and two relative phases of the $\theta_{\eta}$.

In many cases, $\phi_w$ is expected to be large, leading to the
possibility of large $CP$ violation.  Thus for decays like $B \to
K^*\,\rho$ or $B_s \to \rho\,\varphi$, $\phi_w=\gamma$ and for $B \to
\rho\,\rho$ or $B_s \to K^*\,\rho$, $\phi_w=\beta+\gamma$.  For the
parameters $K_5$ and $K_6$ as well as $L_4$, the $CP$ violation is
given by
\begin{equation}
Im[A_{\eta}A_{\eta^{\prime}}^*]
- Im[{\bar A_{\eta}}{\bar A_{\eta^{\prime}}^*}]
= 2\sin\phi_w
\left[
  P_{\eta}T_{\eta^{\prime}}
    \cos(\theta_{\eta}-\theta_{\eta^{\prime}}+\delta_{\eta})
- P_{\eta^{\prime}}T_{\eta}
    \cos(\theta_{\eta}-\theta_{\eta^{\prime}}-\delta_{\eta^{\prime}})
\right].
\end{equation}
Assuming the strong phases are not very large, the major requirement
for a large effect in the above $CP$ asymmetry quantities is that
$P_{\eta}/T_{\eta}$ be quite different from
$P_{\eta^{\prime}}/T_{\eta^{\prime}}$ for $\eta\not=\eta^{\prime}$.
For the case of $K_1$ to $K_3$, the $CP$ violation is given by
\begin{equation}
\frac{|A_{\eta}|^2-|\bar A_{\eta}|^2}
     {|A_{\eta}|^2+|\bar A_{\eta}|^2}
=\frac{-2 T_{\eta}P_{\eta} \sin\phi_w \sin\delta_{\eta}}
      {T_{\eta}^2 + P_{\eta}^2 
       + 2T_{\eta}P_{\eta} \cos\phi_w \cos\delta_{\eta}}
\end{equation}
requiring as noted a significant value for $\sin\delta_{\eta}$.  For
the parameters $K_4$, $L_5$, and $L_6$, the $CP$ violation is measured
by
\begin{equation}
Re[A_{\eta}A_{\eta^{\prime}}^*]
- Re[{\bar A_{\eta}}{\bar A_{\eta^{\prime}}^*}]
= -2\sin\phi_w
\left[
  P_{\eta}T_{\eta^{\prime}}
    \sin(\theta_{\eta}-\theta_{\eta^{\prime}}+\delta_{\eta})
- P_{\eta^{\prime}}T_{\eta}
    \sin(\theta_{\eta}-\theta_{\eta^{\prime}}-\delta_{\eta^{\prime}})
\right].
\end{equation}
These require a significant value of
$\sin(\theta_{\eta}-\theta_{\eta^{\prime}})$ or
$\sin(\delta_{\eta}-\delta_{\eta^{\prime}})$ to have a large effect.


To summarize this paper, we have discussed in a model independent way
the observables in the decay of a $B$ meson to two vector mesons and
the relations among them.  An alternative way of getting the
differential angular distribution is provided in the Appendix.  We
have also explicitly shown how one can determine $L_{4,5,6}$ defined
in the text by measuring the polarization of one of the decay products
in the final state.

This research is supported by the Department of Energy under Grant
No. DE-FG02-91ER40682.


\newpage
\begin{appendix}
\section*{Correlations of Polarizations}

Analyzing the correlations of polarization vectors appearing in the
decay rate provides a way of understanding why only six of the nine
amplitude bilinears show up in the differential cross section.  The
polarization state of a spin 1 particle is described by the density
matrix \cite{L55} which can be written as a sum of a scalar, vector,
and traceless second-rank tensor.  With $a=1,2$ for $V_1$ and $V_2$
mesons, these are
\begin{eqnarray}
\label{tensors}
{\bf 1}_{ij} &=& \delta_{ij}, \qquad
                 {\rm scalar}; \nonumber \\
{\bf S}^a_i &=& \frac1{2i} \varepsilon_{ijk} \epsilon^a_j
                 \epsilon^{a*}_k, \qquad
                 {\rm vector}; \\
{\bf T}^a_{ij} &=& \frac1{2} 
                 \left( \epsilon^a_i \epsilon^{a*}_j
                    + \epsilon^{a*}_i \epsilon^a_j
                    - \frac2{3} \epsilon \cdot \epsilon^* \delta_{ij}
                 \right), \qquad
                 {\rm tensor}. \nonumber
\end{eqnarray}
Therefore, we have
\begin{equation}
\label{decompose}
\epsilon^a_i \epsilon^{a*}_j = 
  {\bf T}^a_{ij} + i\varepsilon_{ijk} {\bf S}_k + \frac1{3}{\bf 1}_{ij},
\end{equation}
provided that the polarization vector $\vec{\epsilon^a}$ are
normalized to $1$.

From Eq.~(\ref{ampl}), one can get the polarization vector
correlations for each of the amplitude bilinears.  After
simplification, we obtain the following results: \linebreak For
$|A_0|^2$:
\begin{eqnarray}
\label{X1}
X_1 &=&
\epsilon^{1*L} \epsilon^{1L} \epsilon^{2*L} \epsilon^{2L} \\
&=& \left(\hat{\bf p}\cdot{\bf T}^1\cdot\hat{\bf p}\right)
    \left(\hat{\bf p}\cdot{\bf T}^2\cdot\hat{\bf p}\right)
   + \frac1{3}\hat{\bf p}\cdot\left({\bf T}^1
        + {\bf T}^2\right)\cdot\hat{\bf p}
   + \frac1{9}. \nonumber
\end{eqnarray}
For $|A_{\|}|^2$:
\begin{eqnarray}
\label{X2}
2X_2 &=&
\vec{\epsilon^{1*T}} \cdot \vec{\epsilon^{2*T}}
  \vec{\epsilon^{1T}} \cdot \vec{\epsilon^{2T}} \\
&=& Tr[{\bf T}^1\cdot{\bf T}^2]
   + \left(\hat{\bf p}\cdot{\bf T}^1\cdot\hat{\bf p}\right)
     \left(\hat{\bf p}\cdot{\bf T}^2\cdot\hat{\bf p}\right)
   - \frac1{3}\hat{\bf p}\cdot\left({\bf T}^1
        + {\bf T}^2\right)\cdot\hat{\bf p}
   \nonumber \\
&&
   - 2 \hat{\bf p}\cdot{\bf T}^1\cdot{\bf T}^2\cdot\hat{\bf p}
   + \frac2{9}
   - 2 \left(\hat{\bf p}\cdot{\bf S}^1\right)
       \left(\hat{\bf p}\cdot{\bf S}^2\right)
   + {\bf S}^1 \cdot {\bf S}^2. \nonumber
\end{eqnarray}
For $|A_{\perp}|^2$:
\begin{eqnarray}
\label{X3}
2X_3 &=&
\vec{\epsilon^{1*}}\times\vec{\epsilon^{2*}}\cdot\hat{\bf p}\;
  \vec{\epsilon^{1}}\times\vec{\epsilon^{2}}\cdot\hat{\bf p} \\
&=& \varepsilon_{ijk}\varepsilon_{lmn}{\bf T}^1_{il}{\bf T}^2_{jm}
       {\bf p}_k {\bf p}_n
    - \frac1{3}\hat{\bf p}\cdot\left({\bf T}^1
         + {\bf T}^2\right)\cdot\hat{\bf p}
    + \frac2{9}. \nonumber
\end{eqnarray}
For $A_0 A_{\|}^*$: (apart from an overall minus sign)
\begin{eqnarray}
\qquad \epsilon^{1*L} \epsilon^{2*L} 
       \vec{\epsilon^{1T}} \cdot \vec{\epsilon^{2T}}
&=& \hat{\bf p}\cdot{\bf T}^1\cdot{\bf T}^2\cdot\hat{\bf p}
   - \left(\hat{\bf p}\cdot{\bf T}^1\cdot\hat{\bf p}\right)
     \left(\hat{\bf p}\cdot{\bf T}^2\cdot\hat{\bf p}\right) 
   -i\hat{\bf p}\cdot\left({\bf T}^2\cdot\hat{\bf p}\right)
       \times{\bf S}^1
     \nonumber \\
&&
   -i\hat{\bf p}\cdot\left({\bf T}^1\cdot\hat{\bf p}\right)
       \times{\bf S}^2
   - {\bf S}^1 \cdot {\bf S}^2
   + \left(\hat{\bf p}\cdot{\bf S}^1\right)
       \left(\hat{\bf p}\cdot{\bf S}^2\right).
\nonumber
\end{eqnarray}
For $A_0^* A_{\|}$: (apart from an overall minus sign)
\begin{eqnarray}
\qquad \epsilon^{1L} \epsilon^{2L} 
       \vec{\epsilon^{1*T}} \cdot \vec{\epsilon^{2*T}}
&=& \hat{\bf p}\cdot{\bf T}^1\cdot{\bf T}^2\cdot\hat{\bf p}
   - \left(\hat{\bf p}\cdot{\bf T}^1\cdot\hat{\bf p}\right)
     \left(\hat{\bf p}\cdot{\bf T}^2\cdot\hat{\bf p}\right) 
   +i\hat{\bf p}\cdot\left({\bf T}^2\cdot\hat{\bf p}\right)
       \times{\bf S}^1
     \nonumber \\
&&
   +i\hat{\bf p}\cdot\left({\bf T}^1\cdot\hat{\bf p}\right)
       \times{\bf S}^2
   - {\bf S}^1 \cdot {\bf S}^2
   + \left(\hat{\bf p}\cdot{\bf S}^1\right)
       \left(\hat{\bf p}\cdot{\bf S}^2\right).
\nonumber
\end{eqnarray}
So the net result for $A_0 A_{\|}^*$ and $A_0^* A_{\|}$ is
\begin{eqnarray}
\label{X4}
X_4&=&
\sqrt{2} \, \left[
       \left(\hat{\bf p}\cdot{\bf T}^1\cdot\hat{\bf p}\right)
       \left(\hat{\bf p}\cdot{\bf T}^2\cdot\hat{\bf p}\right)
      - \hat{\bf p}\cdot{\bf T}^1\cdot{\bf T}^2\cdot\hat{\bf p}
      - {\bf S}^1\cdot{\bf S}^2
      + \left({\bf S}^1\cdot\hat{\bf p}\right)
        \left({\bf S}^2\cdot\hat{\bf p}\right)
     \right],
\nonumber \\
Y_4&=&
\sqrt{2} \, \left[
      \hat{\bf p}\cdot\left({\bf T}^1\cdot\hat{\bf p}\right)
         \times{\bf S}^2
    + \hat{\bf p}\cdot\left({\bf T}^2\cdot\hat{\bf p}\right)
         \times{\bf S}^1
     \right].
\end{eqnarray}
For $A_0 A_{\perp}^*$: (apart from an $i$)
\begin{eqnarray}
\qquad \vec{\epsilon^{1}}\times\vec{\epsilon^{2}}\cdot\hat{\bf p}
       \, \epsilon^{1*L} \epsilon^{2*L}
&=& \hat{\bf p}\cdot\left({\bf T}^1\cdot\hat{\bf p}\right)\times
    \left({\bf T}^2\cdot\hat{\bf p}\right)
   - \hat{\bf p}\cdot{\bf S}^1\times{\bf S}^2
   -i\hat{\bf p}\cdot{\bf T}^2\cdot{\bf S}^1
   \nonumber \\
&&
   +i\hat{\bf p}\cdot{\bf T}^1\cdot{\bf S}^2
   -i\left(\hat{\bf p}\cdot{\bf T}^1\cdot\hat{\bf p}\right)
     \left(\hat{\bf p}\cdot{\bf S}^2\right)
   +i\left(\hat{\bf p}\cdot{\bf T}^2\cdot\hat{\bf p}\right)
     \left(\hat{\bf p}\cdot{\bf S}^1\right).
\nonumber
\end{eqnarray}
With a similar expression for $A_0^* A_{\perp}$, we obtain
\begin{eqnarray}
\label{X5}
X_5&=&
\sqrt{2} \,\left[
      \hat{\bf p}\cdot{\bf T}^2\cdot{\bf S}^1
    - \hat{\bf p}\cdot{\bf T}^1\cdot{\bf S}^2
    + \left(\hat{\bf p}\cdot{\bf T}^1\cdot\hat{\bf p}\right)
      \left({\bf S}^2\cdot\hat{\bf p}\right)
    - \left(\hat{\bf p}\cdot{\bf T}^2\cdot\hat{\bf p}\right)
      \left({\bf S}^1\cdot\hat{\bf p}\right)
     \right],
\nonumber \\
Y_5&=&
\sqrt{2} \, \left[
       \hat{\bf p}\cdot\left({\bf T}^1\cdot\hat{\bf p}\right)\times
       \left({\bf T}^2\cdot\hat{\bf p}\right)
     - \hat{\bf p}\cdot{\bf S}^1\times{\bf S}^2
     \right].
\end{eqnarray}
For $A_{\|} A_{\perp}^*$:(apart from an $i$)
\begin{eqnarray}
\qquad \vec{\epsilon^{1}}\times\vec{\epsilon^{2}}\cdot\hat{\bf p}
       \, \vec{\epsilon^{1*T}}\cdot\vec{\epsilon^{2*T}}
&=& -\hat{\bf p}\cdot\left({\bf T}^1\cdot\hat{\bf p}\right)\times
    \left({\bf T}^2\cdot\hat{\bf p}\right)
   + \varepsilon_{ijk}\left({\bf T}^1\cdot{\bf T}^2\right)_{ij}{\bf p}_k
   + \frac2{3}i\hat{\bf p}\cdot{\bf S}^1 \nonumber \\
&&
   - \frac2{3}i\hat{\bf p}\cdot{\bf S}^2
   +i\left(\hat{\bf p}\cdot{\bf T}^1\cdot\hat{\bf p}\right)
     \left(\hat{\bf p}\cdot{\bf S}^2\right)
   -i\left(\hat{\bf p}\cdot{\bf T}^2\cdot\hat{\bf p}\right)
     \left(\hat{\bf p}\cdot{\bf S}^1\right).
\nonumber
\end{eqnarray}
With a similar expression for $A_{\|}^* A_{\perp}$, we obtain
\begin{eqnarray}
\label{X6}
X_6&=&
     \frac2{3}\hat{\bf p}\cdot\left({\bf S}^2-{\bf S}^1\right)
   + \left(\hat{\bf p}\cdot{\bf T}^2\cdot\hat{\bf p}\right)
     \left({\bf S}^1\cdot{\bf p}\right)
   - \left(\hat{\bf p}\cdot{\bf T}^1\cdot\hat{\bf p}\right)
     \left({\bf S}^2\cdot\hat{\bf p}\right),
\nonumber \\
Y_6&=&
  \varepsilon_{ijk}\left({\bf T}^1\cdot{\bf T}^2\right)_{ij}{\bf p}_k
   - \hat{\bf p}\cdot\left({\bf T}^1\cdot\hat{\bf p}\right)\times
     \left({\bf T}^2\cdot\hat{\bf p}\right).
\end{eqnarray}

Notice that the observables $Y_4$, $X_5$, and $X_6$ are linear in
${\bf S}^1$ or ${\bf S}^2$.  As a result, as discussed in the text,
they cannot be detected via the angular distribution of the decays of
$V_1$ and $V_2$.  However, in principle they may be observed in more
complicated decays, or in decays like $B \to K^*(\to PP) J/\Psi(\to
l^+ l^-)$ by measuring the spin of one of the leptons.  In particular
$X_6$ contains $\hat{\bf p} \cdot {\bf S}^2$ and $\hat{\bf p} \cdot
{\bf S}^1$ and so could be observed by measuring the polarization of
one of the mesons from the decay of $V_1$ or $V_2$ without observing
the other decay.

The above results can be directly applied to the decays $B \to
V_1\,(\to P\,P) \, V_2\,(\to P\,P)$ to obtain the angular
distribution, Eq.~(\ref{BtoVPPVPP}), which is uniquely determined by
the tensor polarizations of the vector mesons.  The angular
distributions of $B \to V_1\,(\to P\,P) \, V_2\,(\to l^+\,l^-)$ and $B
\to V_1\,(\to P\,\gamma) \, V_2\,(\to P\,\gamma)$ can be obtained by
taking into account that the lepton or photon motions must be
perpendicular to the parent particle polarization vector and all
possible spins are summed over.

For the case of decay into pseudoscalars $B \to V_1\,(\to
P_1\,P_1^{\prime}) \, V_2\,(\to P_2\,P_2^{\prime})$, one can go
directly from $X_1$-$X_4$, $Y_5$, and $Y_6$ to the angular
distribution Eq.~(\ref{BtoVPPVPP}).  The polarization vectors of $V_2$
directly convert to the outgoing relative momentum vectors of the
pseudoscalars.  Choosing $\hat{\bf p}=(0,0,1)$, the momentum of $P_1$,
$k_1=(\sin\theta_1,0,\cos\theta_1)$, the momentum of $P_2$,
$k_2=(\sin\theta_2\cos\phi,\sin\theta_2\sin\phi,\cos\theta_2)$, we
have
\begin{eqnarray}
\label{PPPP}
\vec{\epsilon^1} &\to& (\sin\theta_1,0,\cos\theta_1),
\nonumber\\
\vec{\epsilon^2} &\to& (\sin\theta_2 \cos\phi,
                      \sin\theta_2 \sin\phi,\cos\theta_2),
\nonumber \\
{\bf T}^1 &\to& 
  \left(
    \begin{array}{ccc}
      \sin^2\theta_1 - \frac1{3} & 0 & \sin\theta_1 \cos\theta_1 \\
      0 & -\frac1{3} & 0 \\
      \sin\theta_1 \cos\theta_1 & 0 & \cos^2\theta_1 - \frac1{3}
    \end{array}
  \right),
\nonumber \\
{\bf T}^2 &\to& 
  \left(
    \begin{array}{ccc}
      \sin^2\theta_2 \cos^2\phi - \frac1{3} & 
         \sin^2\theta_2 \sin\phi \cos\phi & 
         \sin\theta_2 \cos\theta_2 \cos\phi \\
      \sin^2\theta_2 \sin\phi \cos\phi &
         \sin^2\theta_2 \sin^2\phi - \frac1{3} &
         \sin\theta_2 \cos\theta_2 \sin\phi \\
      \sin\theta_2 \cos\theta_2 \cos\phi &
         \sin\theta_2 \cos\theta_2 \sin\phi &
         \cos^2\theta_2-\frac1{3}
    \end{array}
  \right).
\end{eqnarray}
Putting Eq.~(\ref{PPPP}) into Eqs.~(\ref{X1}-\ref{X6}) one can
immediately get Eq.~(\ref{BtoVPPVPP}).  Notice that terms involving
${\bf S}^a$ make no contribution to the result.

For the case of the decay $B \to V_1\,(\to P_1\,P_1^{\prime}) \, V_2\,(\to
l^+\,l^-)$, suppose we observe that $l^-$ is a right-handed particle
and comes out in the direction
$k_2=(\sin\theta_2\cos\phi,\sin\theta_2\sin\phi,\cos\theta_2)$ with
$\hat{\bf p}=(0,0,1)$ and the momentum of $P_1$,
$k_1=(\sin\theta_1,0,\cos\theta_1)$, we
have instead
\begin{eqnarray}
\label{PPLL}
\vec{\epsilon^1} &\to& (\sin\theta_1,0,\cos\theta_1),
\nonumber\\
\vec{\epsilon^2} &\to& \frac1{\sqrt{2}}
  (\cos\theta_2 \cos\phi - i \sin\phi,
   \cos\theta_2 \sin\phi + i \cos\phi, -\sin\theta_2),
\nonumber \\
{\bf T}^1 &\to& 
  \left(
    \begin{array}{ccc}
      \sin^2\theta_1 - \frac1{3} & 0 & \sin\theta_1 \cos\theta_1 \\
      0 & -\frac1{3} & 0 \\
      \sin\theta_1 \cos\theta_1 & 0 & \cos^2\theta_1 - \frac1{3}
    \end{array}
  \right),
\nonumber \\
{\bf S}^2 &\to&
  \frac1{2}
  \left(
    \begin{array}{c}
      \sin\theta_2 \cos\phi \\
      \sin\theta_2 \sin\phi \\
      \cos\theta_2
    \end{array}
  \right)
\nonumber \\
{\bf T}^2 &\to& 
  \frac1{2}
  \left(
    \begin{array}{ccc}
      \cos^2\theta_2 \cos^2\phi + \sin^2\phi - \frac2{3} & 
         - \sin^2\theta_2 \sin\phi \cos\phi & 
         - \sin\theta_2 \cos\theta_2 \cos\phi \\
      - \sin^2\theta_2 \sin\phi \cos\phi &
         \cos^2\theta_2 \sin^2\phi + \cos^2\phi - \frac2{3} &
         - \sin\theta_2 \cos\theta_2 \sin\phi \\
      - \sin\theta_2 \cos\theta_2 \cos\phi &
         - \sin\theta_2 \cos\theta_2 \sin\phi &
         \sin^2\theta_2-\frac2{3}
    \end{array}
  \right).
\end{eqnarray}
Putting Eq.(\ref{PPLL}) into Eqs.~(\ref{X1}-\ref{X6}) one can
immediately get the differential angular distribution for the decay $B
\to V_1\,(\to P_1\,P_2) \, V_2\,(\to l^+\,l^-)$ with a right-handed
$l^-$ coming out in the final state:
\begin{eqnarray}
\label{BtoPPLL}
\frac1{\Gamma_0}\frac{d^3\Gamma}{d\cos\theta_1 d\cos\theta_2 d\phi}
&=&
\frac{9}{16\pi\Gamma_0}
\biggl \{
  K_1 \cos^2\theta_1 \sin^2\theta_2
+ \frac{K_2}{2}
     \left(
       \sin^2\theta_1 \cos^2\theta_2 \cos^2\phi 
       + \sin^2\theta_1 \sin^2\phi
     \right)
\biggr. \nonumber \\
&& 
+ \frac{K_3}{2}
     \left(
       \sin^2\theta_1 \cos^2\theta_2 \sin^2\phi 
       + \sin^2\theta_1 \cos^2\phi
     \right)
+ \frac{K_4}{2 \sqrt{2}}
       \sin 2\theta_1 \sin 2\theta_2 \cos\phi \nonumber \\
&&
- \frac{K_5}{2 \sqrt{2}}
       \sin 2\theta_1 \sin 2\theta_2 \sin \phi
- \frac{K_6}{2}
       \sin^2\theta_1 \sin^2\theta_2 \sin 2\phi \nonumber \\
&&
\biggl.
+ \frac{L_4}{\sqrt{2}}
       \sin 2\theta_1 \sin \theta_2 \sin\phi
- \frac{L_5}{\sqrt{2}}
       \sin 2\theta_1 \sin \theta_2 \cos\phi
+ \frac{L_6}{2}
     \sin^2\theta_1 \cos\theta_2
\biggr \}
\end{eqnarray}
Notice that terms involving ${\bf S}^1$ do not contribute to the
result.  To obtain the result for the other possible final state with
an left-handed outgoing $l^-$, one only needs to flip the sign of
${\bf S}^2$ and thus the signs of the coefficients of $L_4$, $L_5$,
and $L_6$ (namely, the signs of $Y_4$, $X_5$, and $X_6$).  The muon
polarization is equal to the sum of the terms $L_4$, $L_5$, $L_6$
divided by the sum of the other 6 terms.  For the case of $L_6$ it is
seen that the polarization does not vanish after integrating over
$\theta_1$ and $\phi$ and so the observation can be made without
observing the $V_1$ decay.

\end{appendix}


{\tighten

}

\end{document}